\documentclass[aoas,preprint]{imsart}

\RequirePackage[OT1]{fontenc}
\RequirePackage{amsthm,amsmath}
\RequirePackage{natbib}
\RequirePackage[colorlinks,citecolor=blue,urlcolor=blue]{hyperref}

\usepackage{multirow}
\usepackage{amssymb}
\usepackage{graphicx}
\usepackage{graphicx}
\usepackage{comment}
\usepackage{mathrsfs}
\usepackage[colorinlistoftodos]{todonotes}
\usepackage{rotating}
\usepackage{lscape}
\usepackage{epstopdf}
\usepackage{natbib}

\usepackage{lipsum}
\def\undertilde#1{\mathord{\vtop{\ialign{##\crcr
$\hfil\displaystyle{#1}\hfil$\crcr\noalign{\kern1.5pt\nointerlineskip}
$\hfil\tilde{}\hfil$\crcr\noalign{\kern1.5pt}}}}}

\newcommand{\boldeta}{\boldsymbol{\eta}}
\newcommand{\boldmu}{\boldsymbol{\mu}}
\newcommand{\boldtheta}{\boldsymbol{\theta}}
\newcommand{\boldzeta}{\boldsymbol{\zeta}}
\newcommand{\boldGamma}{\boldsymbol{\Gamma}}

\newcommand{\boldSigma}{\boldsymbol{\Sigma}}
\newcommand{\boldS}{\boldsymbol{S}}

\begin{document}
\begin{frontmatter}
\title{The Autoregressive Structural Model for analyzing longitudinal health data of an aging population in China}

\begin{aug}
\author{\fnms{Yazhuo} \snm{Deng}\thanksref{t1, m1}\ead[label=e1]{dengyazhuo@gmail.com}},
\author{\fnms{David R.} \snm{Paul}\thanksref{t1, m1}\ead[label=e2]{dpaul@uidaho.edu}},
\and
\author{\fnms{Audrey Qiuyan} \snm{Fu}\thanksref{t1, m1}\ead[label=e3]{audreyf@uidaho.edu}}

\thankstext{t1}{Data used in the preparation of this article were obtained from the China Health and Retirement Longitudinal Study. As such, the investigators within the CHARLS contributed to the design
and implementation of CHARLS and/or provided data but did not participate in analysis or writing of this report.}
\runauthor{Deng et al.}
\runtitle{Autoregressive Structural Model}

\affiliation{University of Idaho\thanksmark{m1}}

\address{Yazhuo Deng\\
Department of Movement Sciences\\
Department of Statistical Science\\
David R. Paul\\
Department of Movement sciences\\
Audrey Q. Fu\\
Department of Statistical Science\\
University of Idaho\\
Moscow, Idaho 83844\\
USA\\
\printead{e1}\\
\phantom{E-mail:\ }\printead*{e2}\\
\phantom{E-mail:\ }\printead*{e3}}

\end{aug}

\begin{abstract}
We seek to elucidate the impact of social activity, physical activity and functional health status (factors) on depressive symptoms (outcome) in the China Health and Retirement Longitudinal Study (CHARLS), a multi-year study of aging involving 20,000 participants 45 years of age and older.
Although a variety of statistical methods are available for analyzing longitudinal data, modeling the dynamics within a complex system remains a difficult methodological challenge. We develop an Autoregressive Structural Model (ASM) to examine these factors on depressive symptoms while accounting for temporal dependence. The ASM builds on the structural equation model and also consists of two components: a measurement model that connects observations to latent factors, and a structural model that delineates the mechanism among latent factors. Our ASM further incorporates autoregressive dependence into both components for repeated measurements.  The results from applying the ASM to the CHARLS data indicate that social and physical activity independently and consistently mitigated depressive symptoms over the course of five years, by mediating through functional health status.
\end{abstract}

\begin{keyword}
\kwd{Autoregressive structural model}
\kwd{Template structure}
\kwd{Autoregressive model}
\kwd{Structural equation modeling}
\end{keyword}

\end{frontmatter}

\section{Introduction}
\label{introduction}

The aging population in China is growing rapidly: it is estimated that the proportion of the population aged 60 or older will increase from 10\% in 2000 to about 30\% in 2050 [\cite{banister2012population}]. As the burden of ischaemic heart disease, high systolic blood pressure and mental health disorders grows along with the economic development in China, the nature of health problem will shift from infectious to chronic illnesses [\cite{zhou2019mortality}]. The China Health and Retirement Longitudinal Study (CHARLS) [\cite{zhao2012cohort}] is an ongoing longitudinal study of Chinese adults 45 years of age or older that reside in administrative villages in rural areas and neighborhoods in urban areas across China. Approximately 20,000 respondents from 150 counties in 20 provinces participated in the baseline and follow-up surveys designed to assess their demographic characteristics, and socioeconomic and health change. This study provides a high-quality public micro-database for scientific and policy research on aging-related issues. 

Late-life depression may be linked to the deterioration of many medical conditions, and subsequently negatively impacts the quality of life [\cite{fiske2009depression}].  Unfortunately, it is often neglected in the elderly [\cite{xu2016decomposing}]. Social activity and physical activity may have a direct effect on the elderly's depressive symptoms, but the underlying mechanism by which multiple socio-behavioral factors simultaneously affect depressive symptoms over time can be complex [\cite{fried2004social}].  We and other groups have also demonstrated that functional health status may act as a mediator in these pathways [\cite{fried2004social}; \cite{deng2018relationships}]; a mediator is an intermediate factor through which another factor influences an outcome (depressive symptoms here) [\cite{mackinnon2008introduction}].  However, these existing studies do not investigate the complex dynamics of these interrelationships.

A variety of methods under the framework of the Structural Equation Modeling (SEM) have been developed to model the dynamics of a complex system [see \cite{little2013longitudinal}; \cite{mcardle2014longitudinal}; \cite{ferrer2003alternative}]. The standard approaches include the latent growth models, which use the latent variables of the intercept and slope to describe the growth trajectory of the outcome variable [\cite{grimm2016growth}], and the autoregressive models, such as the simplex model that measures the temporal change of repeated measurements [\cite{marsh1993stability}]. The autoregressive latent trajectory model developed by \cite{bollen2004autoregressive} combines the desirable features of the two approaches above to describe the individual growth trajectory of the outcome while accounting for the temporal dependence, but it is still unable to account for any factor. Similarly, although the dynamic structural equation models  [\cite{asparouhov2018dynamic}] combines time-series models with SEM, it is designed for intensive longitudinal data with a large number of time points of a single outcome, and also cannot account for factors. 

Therefore, multivariate extensions of autoregressive models are useful for studying multidimensional structural relationships over time. A classical model is the cross-lagged panel model [\cite{mayer1986cross}], which estimates the regression coefficients between two repeatedly measured variables (a factor and an outcome) and their autoregressive coefficients over time. Extending this model, the Autoregressive Mediation Model (AMM) adds a third, time-lagged variable (the second factor) for longitudinal mediation analysis [\cite{cole2003testing, mackinnon2008introduction}]. Using longitudinal data to test hypotheses on mediation allows us to control for the effects of earlier time points, hence minimizing bias in the estimated structural relationships [\cite{selig2009mediation}]. However, the AMM can consider only two factors and an outcome.  Therefore, it is desirable to develop statistical models that simultaneously assess multiple structural relationships and mediations over time. 

In this paper, we generalize the AMM and propose the Autoregressive Structural Model (ASM) to capture the dynamics of a complex system.  The dynamics are described by an autoregressive model, and the mechanism at each time point is described by an SEM.  The ASM can be visualized as a direct acyclic graph (DAG): the variables of interest (such as factors and outcomes) are the nodes and appear at multiple time points, and directed edges represent the mechanistic relationships within a time point as well as autoregressive dependence between time points. We assume the SEM at any time to have an identical structure, although the coefficients may differ.  We term this SEM the \textit{template}, similar to the terminology used in the temporal graphical model developed by \cite{koller2009probabilistic}. In the next section, we describe the details of the ASM and related issues, such as covariate adjustment and additional assumptions on the parameters. We apply the ASM to the CHARLS in Section~\ref{application}, and discuss the implications and caveats in Section~\ref{discussion}.

\section{Statistical models}
\label{model}

\subsection{The Structural Equation Modeling} 
The SEM focuses on the modeling of variances and covariances of multivariate variables [\cite{Bollen1989Structural}].  In the SEM framework, the relationships between latent and observed variables are expressed using a measurement equation and a structural equation.  Consider $p$ observed variables (factors and outcomes) and $q$ latent variables.  The measurement equation can be written as
\begin{equation}\label{eqn:measure}
\mathbf{y}=\boldsymbol{\mu+ \Lambda \eta + \epsilon},
\end{equation}
where $\mathbf{y}$ is a random vector of the $p$ observed variables, $\boldsymbol{\mu}$ an intercept vector, $\mathbf{\Lambda}$ a $p \times q$ factor loading matrix, $\boldsymbol{\eta}$ a vector of the $q$ latent variables, and $\boldsymbol{\epsilon}$ a random vector of residuals. Additionally, we define $\mathbf{\Theta}$ as the covariance matrix of $\boldsymbol{\epsilon}$.

The structural equation represents the relationships among latent variables as 
\begin{equation}\label{eqn:structural}
\boldsymbol{\eta = \Gamma \eta + \zeta },
\end{equation}
where $\boldsymbol{\Gamma}$ is a $q \times q$ coefficient matrix that indicates the structural relationships among latent variables, and $\boldsymbol{\zeta}$ also a random vector of residuals.  We denote the covariance matrix of $\boldsymbol{\zeta}$ by $\mathbf{\Psi}$.

We can rewrite the structural equation (Equation~\ref{eqn:structural}) as 
\begin{equation*}
\boldsymbol{\eta=\mathbf{(I-\boldsymbol{\Gamma})}^{-1}\zeta},
\end{equation*}
and plug this expression in the measurement equation (Equation~\ref{eqn:measure}) to have
\begin{equation*}
\mathbf{y= \boldmu + \Lambda \mathbf{(I-\boldsymbol{\Gamma})}^{-1}\boldsymbol{\zeta + \epsilon} }.
\end{equation*}
Then the covariance structure can be formulated as 
\begin{align*}
\boldsymbol{\Sigma} &= \mathbb E\{\mathbf{(y-\boldmu) (y-\boldmu)}^{T}\} = \mathbb E\{(\mathbf{\Lambda \mathbf{(I-\boldsymbol{\Gamma})}^{-1}\boldsymbol{\zeta + \epsilon }})(\mathbf{\Lambda \mathbf{(I-\boldsymbol{\Gamma})}^{-1}\boldsymbol{\zeta + \epsilon }})^T\} \\
&= \boldsymbol{\Lambda}\mathbf{(I-\boldsymbol{\Gamma})^{-1} \Psi ((I-\boldsymbol{\Gamma})^{-1})}^T \mathbf{\Lambda}^T + \mathbf{\Theta}.
\end{align*}
where $(\mathbf{I-\boldsymbol{\Gamma}})$ is a nonsingular matrix.  The parameter vector $\boldsymbol{\theta}$ of the SEM is then
\begin{align*}
\boldsymbol{\theta} = \{\boldmu, \mathbf{\Lambda}, \boldsymbol{\Gamma}, \mathbf{\Psi}, \mathbf{\Theta}\}.
\end{align*}

\subsection{The longitudinal measurement model} 

For longitudinal data collected at $T$ time points, the measurement equation at the $t$th time point is then
\begin{equation}\label{eqn:measure_t}
\mathbf{y}_t=\boldsymbol{\mu}_t +\mathbf{\Lambda}_t \boldsymbol{\eta}_t + \boldsymbol{\epsilon}_t.
\end{equation}

The set of measurement equations across all time points can be stacked up to have 
\begin{equation*}
\begin{bmatrix}
\mathbf{y}_1 \\
\mathbf{y}_2 \\
\vdots  \\
\mathbf{y}_T
\end{bmatrix}
=
\begin{bmatrix}
\boldsymbol{\mu}_1 \\
\boldsymbol{\mu}_2 \\
\vdots  \\
\boldsymbol{\mu}_T
\end{bmatrix}
+
\begin{bmatrix}
\mathbf{\Lambda}_1 & \mathbf{0} & \dots & \mathbf{0}\\
\mathbf{0} & \mathbf{\Lambda}_2 &\dots & \mathbf{0}\\
\vdots & \vdots & \ddots & \vdots \\
\mathbf{0} & \mathbf{0} & \dots & \mathbf{\Lambda}_T \\
\end{bmatrix}
\begin{bmatrix}
\boldsymbol{\eta}_1 \\
\boldsymbol{\eta}_2 \\
\vdots  \\
\boldsymbol{\eta}_T
\end{bmatrix}
+
\begin{bmatrix}
\boldsymbol{\epsilon}_1 \\
\boldsymbol{\epsilon}_2 \\
\vdots  \\
\boldsymbol{\epsilon}_T
\end{bmatrix},
\end{equation*}
which we can concisely represent as
\begin{align}\label{eqn:asmsem}
\mathbf{y=\boldsymbol{\mu +\Lambda \eta + \epsilon}}.
\end{align}

\subsection{The Autoregressive Structural Model (ASM)} 
We propose the ASM that combines the two models above to capture the structural relationships behind longitudinal measurements.  Similar to the SEM, our ASM also contains a measurement equation and a structural equation.
The structural equation of the ASM at the $t$th time can be formulated as  
\begin{align} 
\boldsymbol{\eta}_t &=\sum_{i=1}^{t-1}{\mathbf{\Pi}_{i \rightarrow t}}\boldsymbol{\eta}_i + \mathbf{B}_t \boldsymbol{\eta}_t +\boldsymbol{\zeta} _t,  \label{eqn:ASM} 
\end{align}
where 
$\mathbf{\Pi}_{i \rightarrow t}$ is the $q\times q$ diagonal matrices containing the (higher-order) autoregressive coefficients from the $i$th time point, and $\mathbf{B}_t$ is the $q\times q$ coefficient matrix among the latent variables at the $t$th time.  Across all time points, the autoregressive structural equations can also be stacked up and concisely written as follows:
\begin{align}
\boldsymbol{\eta} &\equiv \boldsymbol{\Gamma \eta} + \boldsymbol{\zeta},  \label{eqn:asmgeneral}
\end{align}
where $\boldeta$ and $\boldzeta$ are both vectors of length $qt$, and $\boldGamma$ a $qt \times qt$ matrix of coefficients.
We assume the same formulation of the structural equation at all the time points, although the coefficient estimates may be different at different time points.  This identical structural equation is the template in our ASM.

For a first-order autoregressive structural equation, we have 
\begin{equation*}
\boldsymbol{\Pi}_{(t-1) \rightarrow t}=
\begin{bmatrix}
\pi_{1,(t-1) \rightarrow t} & 0 & 0 & 0\\
0 & \pi_{2,(t-1) \rightarrow t} & 0 & 0\\
 \vdots & \vdots & \ddots & \vdots \\
0 & 0 & 0 & \pi_{q,(t-1) \rightarrow t}\\
\end{bmatrix},
\end{equation*}
where $\pi_{k,(t-1) \rightarrow t}$, $k=1,\dots,q$, is the autoregressive coefficient of the $k$th latent variable in the template from the $(t-1)$th to $t$th time. 

For an ASM of order 2,  
the coefficient matrix $\boldsymbol{\Gamma}$ is
\begingroup\makeatletter\def\f@size{10}\check@mathfonts
\def\maketag@@@#1{\hbox{\m@th\large\normalfont#1}}
\begin{equation*}
\boldsymbol{\Gamma}
=
\begin{bmatrix}
\mathbf{B}_1 & \mathbf{0} & \mathbf{0} & \mathbf{0} &\dots  & \mathbf{0}& \mathbf{0}& \mathbf{0}& \mathbf{0}\\
\mathbf{\Pi}_{1 \rightarrow 2} & \mathbf{B}_2 & \mathbf{0} & \mathbf{0} &\dots  & \mathbf{0}& \mathbf{0}& \mathbf{0}& \mathbf{0}\\
\mathbf{\Pi}_{1 \rightarrow 3} & \mathbf{\Pi}_{2 \rightarrow 3} & \mathbf{B}_3 & \mathbf{0} &\dots  & \mathbf{0}& \mathbf{0}& \mathbf{0}& \mathbf{0}\\
\mathbf{0} & \mathbf{\Pi}_{2 \rightarrow 4} & \mathbf{\Pi}_{3 \rightarrow 4} & \mathbf{B}_4 &\dots  & \mathbf{0}& \mathbf{0}& \mathbf{0}& \mathbf{0}\\
\\
\vdots &  \vdots &   \vdots &  \vdots &  \ddots  & \vdots & \vdots & \vdots & \vdots\\
\\
\mathbf{0} & \mathbf{0} & \mathbf{0} & \mathbf{0} &\dots  &\mathbf{\Pi}_{(T-3) \rightarrow (T-1)}&\mathbf{\Pi}_{(T-2) \rightarrow (T-1)}&\mathbf{B}_{T-1}& \mathbf{0}\\
\mathbf{0} & \mathbf{0} & \mathbf{0} & \mathbf{0} &\dots  & \mathbf{0}&\mathbf{\Pi}_{(T-2) \rightarrow T}&\mathbf{\Pi}_{(T-1) \rightarrow T}&\mathbf{B}_T\\
\end{bmatrix}.
\end{equation*}\endgroup

\subsection{Covariate adjustment} 

Demographic and socioeconomic covariates may bias parameter estimation. To control for these confounding variable, we include these covariates and rewrite Equation \ref{eqn:ASM} as follows:
\begin{equation}\label{eqn:covariates}
\boldsymbol{\eta}_t = \sum_{i=1}^{t-1}{\mathbf{\Pi}_{i \rightarrow t}}\boldsymbol{\eta}_i + \mathbf{B}_t \boldsymbol{\eta}_t +\mathbf{C}_t \boldsymbol{\eta}_c+\boldsymbol{\zeta} _t, 
\end{equation}
where the vector $ \boldsymbol{\eta}_c$ contains the time-invariant covariates and the matrix of covariate coefficients $\mathbf{C}_t$ contains elements $c_{k,m,t}$, which is the coefficient of the $m$th covariates affecting the $k$th latent variable at the $t$th time. 

\subsection{Inference and model fit}

Since our ASM can be formulated as an SEM (Equations~\ref{eqn:asmsem} and~\ref{eqn:ASM}, even accounting for covariates), we can use the inference methods developed for SEMs for parameter estimation here.  Specifically, we will use the maximum likelihood method described in Chapter 4 of \cite{Bollen1989Structural}. This method minimizes the differences between the sample covariance matrix $\mathbf{S}$ of the observed variables $\mathbf{y}$ and the estimated covariance matrix $\hat{\boldSigma}$
using a discrepancy function $F(\boldS, \hat{\boldSigma})$.
Note that the estimation of $\boldSigma$ involves estimation of the parameter vector $\boldtheta$.  To emphasize such connection, we use the notation $\mathbf{\Sigma(\boldsymbol{\hat \theta})}$ in place of $\hat{\boldSigma}$, where $\hat{\boldtheta}$ is the estimate of $\boldtheta$.
Assuming multivariate normality of $\mathbf{y}$, minimizing the discrepancy function can be obtained by maximizing the likelihood:
\begin{equation} \label{eqn:ml}
F_{\text{ML}}(\mathbf{S},\mathbf{\Sigma(\boldsymbol{\hat \theta})})=\text{tr}(\mathbf{S\Sigma(\boldsymbol{\hat \theta})}^{-1}) + \log |\mathbf{\Sigma(\boldsymbol{\hat \theta})}|-\log |\mathbf{S}|-p,
\end{equation}
where tr is trace and ML stands for maximum likelihood.  The estimate $\hat{\boldtheta}$ is then the maximum likelihood estimate.

To assess the model fit for the ASM, we can also use the multiple metrics designed for SEMs [\cite{hu1999cutoff}]: 
\begin{itemize}
  \item The $\chi^2$ statistic: This is the minimized discrepancy in Equation~\ref{eqn:ml} and follows a $\chi^2$ distribution with the degrees of freedom of the current model.
  \item The comparative fit index (CFI): It compares a model of interest to the null model which assumes zero covariances among the observed variables and is defined as
$$
\text{CFI}=\frac{d_{\text{null}}-d_{\text{specified}}}{d_{\text{null}}},
$$
where $d_{\text{null}}=\chi^2_{\text{null}}-df_{\text{null}}$ and $d_{\text{specified}}=\chi^2_{\text{specified}}-df_{\text{specified}}$, and $df$ indicates the degrees of freedom of the corresponding model. 
  \item The standardized root mean square residual (SRMR): 
$$
\text{SRMR}=\sqrt{(\sum_{j=1}^p\sum_{k=1}^p r^2_{jk})/e},
$$
where $r_{jk}$ is the difference between the observed and estimated correlation between $y_j$ and $y_k$, and $e=p(p+1)/2$, with $p$ being the number of observed variables. 
  \item The root mean square error of approximation (RMSEA) with its 90\% confidence interval (CI).  
The RMSEA is defined as 
$$
\text{RMSEA}=\sqrt{\frac{(\chi^2_{\text{specified}}/df_{\text{specified}})-1}{n}}.
$$
\end{itemize}

CFI, SRMR, and RMSEA with respective values of greater than .90, less than .08, less than .06 suggest a good model fit [\cite{hu1999cutoff}]. The confidence interval of RMSEA should be below 0.06.  

\subsection{Longitudinal measurement invariance}\label{sec:invar} 

A key assumption of longitudinal SEM models (such as those discussed in the Introduction) is measurement invariance of the latent variables over different time points [\cite{little2013longitudinal}; \cite{millsap2012investigating}]. This assumption ensures that for any latent variable $\boldsymbol{\eta}$, the model measures the same effect over time. Here, { we consider three levels of invariance, namely the configural, weak and strong invariance, which give rise to three nested ASMs [\cite{widaman2010factorial}]. The configural invariance requires the same structural relationships across time, and no constraint on parameters is added to the measurement equation (i.e., Equation~\ref{eqn:asmsem}). This is also our ASM without additional constraints, as we use the same template at all time points. The weak invariance requires equality in factor loadings over time: 
\begin{equation}\label{eqn:loadingInv}
\mathbf{\Lambda}_1 = \mathbf{\Lambda}_2 = \dots = \mathbf{\Lambda}_T.
\end{equation}
The strong invariance requires invariant loadings and invariant intercepts:
\begin{equation}\label{eqn:interceptInv}
\boldsymbol{\mu}_1 = \boldsymbol{\mu}_2 = \dots = \boldsymbol{\mu}_T.
\end{equation}
The strong invariance ASM is therefore nested in the weak invariance ASM, which in turn is nested in the configural invariance ASM. If the weak or strong invariance for all the elements is not supported by data, partial invariance may be imposed for a subset of factor loadings or intercepts. 

Testing these different levels of invariance is effectively model selection.  We use the model fit metrics described above to compare and select models. In particular, we can compute the differences in the CFI values (denoted as $\Delta\text{CFI}$) between two models. When $\Delta\text{CFI}<.01$, the two models do not differ significantly.

\section{Application}
\label{application}

\subsection{The CHARLS study}
We use the first three waves (i.e., time points; 2011, 2013 and 2015) of the CHARLS survey data to investigate the dynamic relationships between the social activity (SA), physical activity (PA), and functional health status (FHS), and their impact on depressive symptoms (DS) among Chinese adults of 45 years and older. In the 2011 national baseline study, a representative sample of 17,708 participants from 150 urban districts and rural counties in 28 provinces were recruited using a multistage probability sampling strategy. In 2013 and 2015 follow-up studies, 18,605 and 21,095 respondents participated respectively, including follow-up respondents and newly added ones. Only 8,959 participants responded to the PA survey at least once in the three waves of measurements.  We focus on these participants in our analysis here.
Among them, 4,739 (53\%) are female, 7,837 (88\%) are married and 5,759 (64\%) reside in rural areas. The proportion of illiterate participants is 33\% and 18\% in rural and urban areas, respectively. Detailed sampling procedures and the cohort profile can be found in \cite{zhao2012cohort}.
 
Among the variables of interest, social activity measures the frequency of engaging in social activities (e.g., interacting with friends; going to community club; attending training course; caring for sick or disabled adult; taking part in the community-related organization, etc.) in the month prior to the survey. Physical activity consists of weekly durations of vigorous activity, moderate activity and walking. Functional health status utilizes the 5-item Instrumental Activities of Daily Living (IADLs) to assess the functional limitations in the engagement of essential skills for independent living; higher scores indicate greater difficulties in performing daily living activities. Depressive symptoms are measured using eight items of the Center for Epidemiologic Studies Depression Scale (CES-D). In total, fifteen observed variables constitute the four latent or observed variables of interest.

We consider four latent variables (one for each category) and construct an ASM of order-2 to examine the relationships among these variables.  Our ASM also controls for the time-invariant covariates, including sex, age, rural/urban residency (URB), marital status (MAR), and educational attainment (EDU).

The longitudinal measurement equation $\mathbf{y}_t=\boldsymbol{\mu}_t +\mathbf{\Lambda}_t \boldsymbol{\eta}_t + \boldsymbol{\epsilon}_t$ at time $t$ ($t = 1, 2, 3 $) can be expressed as follows:
\begin{equation*}
\begin{bmatrix}
y_{1,t} \\
y_{2,t} \\
y_{3,t} \\
\vdots  \\
y_{7,t} \\
y_{8,t} \\
\vdots  \\
y_{15,t} 
\end{bmatrix}
=
\begin{bmatrix}
\mu_{1,t} \\
\mu_{2,t} \\
\mu_{3,t} \\
\vdots  \\
\mu_{7,t} \\
\mu_{8,t} \\
\vdots  \\
\mu_{15,t} 
\end{bmatrix}
+
\begin{bmatrix}
\lambda_{1,t} & 0 & 0 & 0\\
0 & \lambda_{2,t} & 0 & 0\\
0 & 0 & \lambda_{3,t} & 0 \\
0 & 0 & \vdots & \vdots \\
0 & 0 & \lambda_{7,t} & 0 \\
0 & 0 & 0 & \lambda_{8,t} \\
\vdots & \vdots \\
0 & 0 & 0 & \lambda_{15,t}
\end{bmatrix}
\begin{bmatrix}
\eta_{SA,t} \\
\eta_{PA,t}  \\
\eta_{FHS,t} \\
\eta_{DS,t} 
\end{bmatrix}
+
\begin{bmatrix}
\epsilon_{1,t} \\
\epsilon_{2,t} \\
\epsilon_{3,t} \\
\vdots  \\
\epsilon_{7,t} \\
\epsilon_{8,t} \\
\vdots  \\
\epsilon_{15,t}
\end{bmatrix},
\end{equation*}
where $y_{1,t}$ is the observed variable for social activity, $y_{2,t}$ for physical activity, $y_{3,t}$ to $y_{7,t}$ for functional health status,  and $y_{8,t}$ to $y_{15,t}$ for depressive symptoms at time $t$.  In addition, $\eta_{SA}$, $\eta_{PA}$, $\eta_{FHS}$ and $\eta_{DS}$ are the corresponding latent variables. The measurement equations for three time points are stacked to form $\mathbf{y}=\boldsymbol{\mu +\Lambda \eta + \epsilon}$, which can be expressed as 
\begin{equation}\label{eqn:measureCHARLS}
\begin{bmatrix}
\mathbf{y}_1 \\
\mathbf{y}_2 \\
\mathbf{y}_3
\end{bmatrix}
=
\begin{bmatrix}
\boldsymbol{\mu}_1 \\
\boldsymbol{\mu}_2 \\
\boldsymbol{\mu}_3
\end{bmatrix}
+
\begin{bmatrix}
\mathbf{\Lambda}_1 & \mathbf{0}  & \mathbf{0}\\
\mathbf{0} & \mathbf{\Lambda}_2  & \mathbf{0}\\
\mathbf{0} & \mathbf{0} & \mathbf{\Lambda}_3 \\
\end{bmatrix}
\begin{bmatrix}
\boldsymbol{\eta}_1 \\
\boldsymbol{\eta}_2 \\
\boldsymbol{\eta}_3
\end{bmatrix}
+
\begin{bmatrix}
\boldsymbol{\epsilon}_1 \\
\boldsymbol{\epsilon}_2 \\
\boldsymbol{\epsilon}_3
\end{bmatrix}.
\end{equation}

In the structural equation, we assume that social and physical activity may influence functional health status and depressive symptoms, and that functional health status may further influence depressive symptoms.  We do not allow the reverse.  We further assume that the covariates may influence all four latent variables.  A diagram of this structural equation model is depicted in Figure \ref{figure:hypothesized}.  Specifically,
\begin{align}\label{eqn:order2auto}
\begin{bmatrix}
\mathbf{\eta}_{SA,t} \\
\mathbf{\eta}_{PA,t} \\
\mathbf{\eta}_{FHS,t} \\
\mathbf{\eta}_{DS,t} 
\end{bmatrix}&=
\begin{bmatrix}
\pi_{1,(t-2) \rightarrow t}& 0 & 0 & 0\\
0 & \pi_{2,(t-2) \rightarrow t} & 0 & 0\\
 0 & 0 & \pi_{3,(t-2) \rightarrow t} & 0 \\
0 & 0 & 0 & \pi_{4,(t-2) \rightarrow t}\\
\end{bmatrix}
\begin{bmatrix}
\mathbf{\eta}_{SA,(t-2)} \\
\mathbf{\eta}_{PA,(t-2)} \\
\mathbf{\eta}_{FHS,(t-2)} \\
\mathbf{\eta}_{DS,(t-2)} 
\end{bmatrix} \\
&+
\begin{bmatrix}
\pi_{1,(t-1) \rightarrow t} & 0 & 0 & 0\\
0 & \pi_{2,(t-1) \rightarrow t} & 0 & 0\\
 0 & 0 & \pi_{3,(t-1) \rightarrow t} & 0 \\
0 & 0 & 0 & \pi_{4,(t-1) \rightarrow t}\\
\end{bmatrix}
\begin{bmatrix}
\mathbf{\eta}_{SA,(t-1)} \\
\mathbf{\eta}_{PA,(t-1)} \\
\mathbf{\eta}_{FHS,(t-1)} \\
\mathbf{\eta}_{DS,(t-1)} 
\end{bmatrix} \notag\\
&+
\begin{bmatrix}
0 & 0 & 0 & 0\\
0 & 0 & 0 & 0\\
 \beta_{13,t} & \beta_{23,t} & 0 & 0 \\
\beta_{14,t} & \beta_{24,t} & \beta_{34,t} & 0\\
\end{bmatrix}
\begin{bmatrix}
\mathbf{\eta}_{SA,t} \\
\mathbf{\eta}_{PA,t} \\
\mathbf{\eta}_{FHS,t} \\
\mathbf{\eta}_{DS,t} 
\end{bmatrix} \notag\\
&+
\begin{bmatrix}
c_{1,1,t} & c_{1,2,t} & c_{1,3,t} & c_{1,4,t} & c_{1,5,t} \\
c_{2,1,t} & c_{2,2,t} & c_{2,3,t} & c_{2,4,t} & c_{2,5,t}\\
c_{3,1,t} & c_{3,2,t} & c_{3,3,t} & c_{3,4,t} & c_{3,5,t}\\
c_{4,1,t} & c_{4,2,t} & c_{4,3,t} & c_{4,4,t} & c_{4,5,t}
\end{bmatrix}
\begin{bmatrix}
\mathbf{\eta}_{SEX} \\
\mathbf{\eta}_{AGE} \\
\mathbf{\eta}_{URB} \\
\mathbf{\eta}_{MAR} \\
\mathbf{\eta}_{EDU} 
\end{bmatrix}+
\begin{bmatrix}
\mathbf{\zeta}_{SA,t} \\
\mathbf{\zeta}_{PA,t} \\
\mathbf{\zeta}_{FHS,t} \\
\mathbf{\zeta}_{DS,t} 
\end{bmatrix} \notag
\end{align}
In vector notation,
\begin{align}
\boldsymbol{\eta}_t = \mathbf{\Pi}_{(t-2) \rightarrow t} \boldsymbol{\eta}_{(t-2)} + \mathbf{\Pi}_{(t-1) \rightarrow t} \boldsymbol{\eta}_{(t-1)} + \mathbf{B}_t \boldsymbol{\eta}_t + \mathbf{C}_t \boldsymbol{\eta}_c +\boldsymbol{\zeta} _t.
\end{align}

This structural model implies that the information on $\eta_{SA,t}$ and $\eta_{PA,t}$ comes only from the corresponding observed variables without measurement errors:
\begin{align}
y_{1,t} = \mu_{1,t} + \eta_{SA,t}, \;\; \text{and} \;\; y_{2,t} = \mu_{2,t} + \eta_{PA,t}.
\end{align}
In other words,
\begin{align}
\lambda_{1} = \lambda_{2} = 1, \;\; \text{and} \;\;  \epsilon_{1,t}=\epsilon_{2,t} =0.
\end{align}

\begin{figure}[h!]
\begin{center}
  \includegraphics[width=1.0\textwidth]{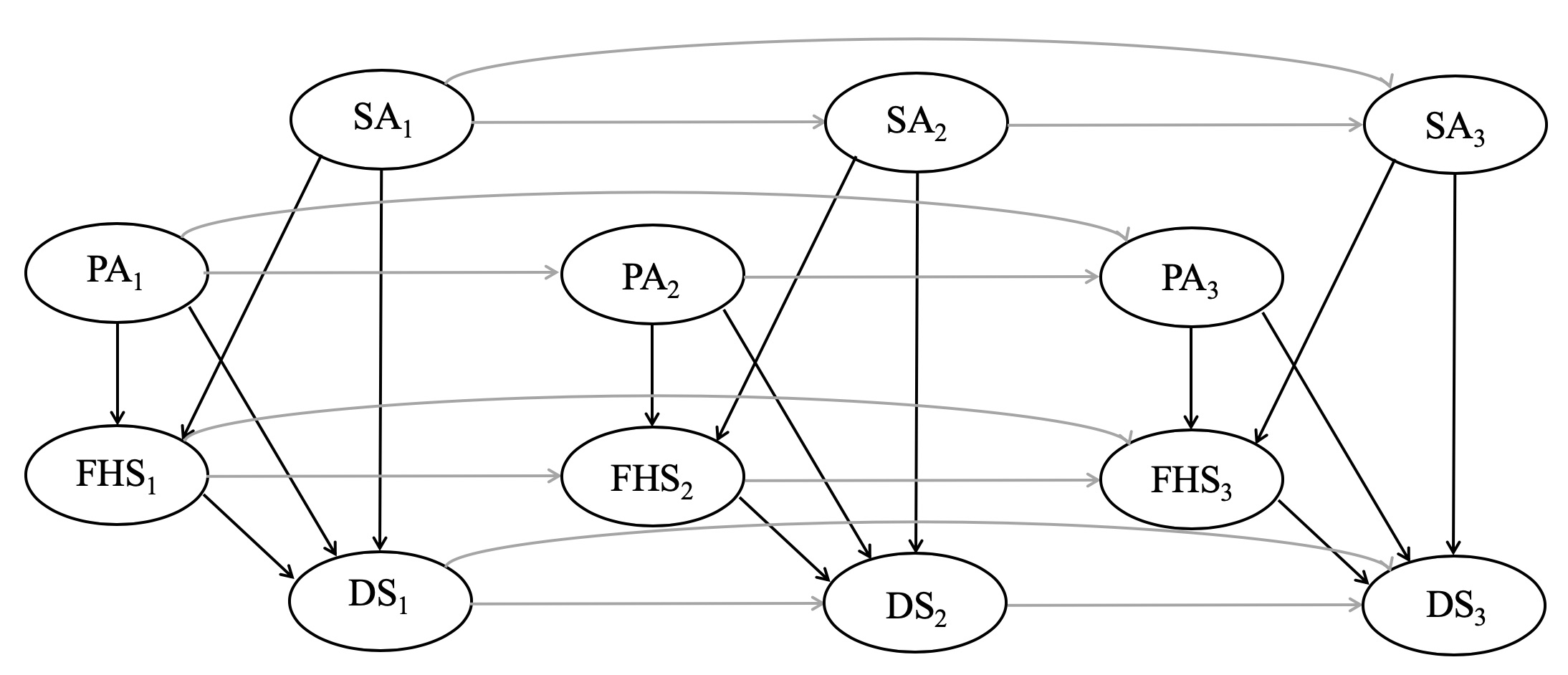}\\
  \caption{The diagram of the structural equation model in our Autoregressive Structural Model (ASM) of order-2 for three time points. SA stands for social activity, PA physical activity, FHS functional health status, and DS depressive symptoms.}
   \label{figure:hypothesized}
\end{center}
\end{figure}

We test different types of measurement invariance as described in Section~\ref{sec:invar}.
Specifically, the metric invariance tests 
\begin{align}
\mathbf{\Lambda}_1 = \mathbf{\Lambda}_2  = \mathbf{\Lambda}_3. 
\end{align}
The scalar invariance further tests 
\begin{align}
\boldsymbol{\mu}_1 = \boldsymbol{\mu}_2 = \boldsymbol{\mu}_3. 
\end{align}
The analyses were conducted using Mplus 8.0 [\cite{muthen2017mplus}] and the Mplus code is in the Supplementary Materials and on Github \url{https://github.com/YazhuoDeng/AutoregressiveStructuralModel}. 

\subsection{Results}

The sample means of the observed variables from the three years are presented in Figure~\ref{fig:compMean} and Table \ref{tab:measureInvar}. Whereas depressive symptoms show a fluctuating pattern between 2011 and 2015, functional health status worsens over time, indicated by increasing mean scores. Furthermore, the means of physical activity decreased from 2011 to 2015, while the means of social activity increased first and then declined. 

Table \ref{table:goodnessoffit} reports the goodness-of-fit statistics for different invariance models described in Section 3.1. The configural invariance ASM yields a good fit. The weak invariance ASM $(\mathbf{\Lambda}_{1} = \mathbf{\Lambda}_{2} = \mathbf{\Lambda}_{3}$ does not different substantially from the configural model $(\Delta \text{CFI}= 0.943 - 0.940 = 0.003)$. In the strong invariance ASM, we further constrain the intercepts of depressive symptoms to be equal across time $(\boldsymbol{\mu}_{i,1} = \boldsymbol{\mu}_{i,2} = \boldsymbol{\mu}_{i,3}$, where $i=8, \cdots, 15$). We do not impose the invariance constraint on the intercepts of functional health status, as the sample mean scores of functional health status noticeably increased from 2011 to 2015.  This (partially) strong invariance model also does not differ substantially from the configural model $(\Delta \text{CFI}= 0.943 - 0.938 = 0.005)$. Therefore, the final model is the (partially) strong invariance model. We summarize the intercepts and factor loadings estimated in the final ASM in Table \ref{tab:measureInvar}.  

\begin{table}[ht]
\centering
\caption{Goodness-of-fit metrics of the ASM with different levels of invariance. The factor loadings of observed variables of functional health status and depressive symptoms were constrained to be invariant in the weak invariance model. The intercepts of observed variables of depressive symptoms were constrained to be invariant in the strong invariance model. df stands for degree of freedom, CFI comparative fit index, RMSEA root mean square error of approximation, and SRMR standardized root mean square residual. }
\bigskip
\renewcommand{\arraystretch}{1.5}
{\begin{tabular}{lccccc}
      \hline
       & $\chi^2$ & \textit{df} & CFI & RMSEA (CI) & SRMR\\
    \hline
      Configural Invariance ASM & 8748.500 & 1044 &0.943 & 0.029 (0.028,0.029) &0.033\\ 
      Weak Invariance ASM & 9135.501 & 1066 &0.940 & 0.029 (0.029,0.030) &0.034\\ 
      Strong Invariance ASM & 9402.683 & 1082 &0.938 & 0.029 (0.029,0.030) &0.035\\  
    \hline
\end{tabular}}
  \label{table:goodnessoffit}
\end{table}

\begin{table}[ht]
\centering
\caption{Observed means ($\bar y$), estimated intercepts ($\mu$) and factor loadings ($\lambda$) in the measurement model of our strong invariance ASM. 
}
\bigskip
\label{tab:measureInvar}
\renewcommand{\arraystretch}{1.5}
{\begin{tabular}{p{3.7cm}ccccccc}
      \hline
      Variable & $\bar y_{1}$ & $\bar y_{2}$ & $\bar y_{3}$ & $\mathbf{\mu}_{1}$ & $\mathbf{\mu}_{2}$ & $\mathbf{\mu}_{3}$ & $\mathbf{\lambda}$ \\
    \hline
      \textit{Social activity}\\
      \hspace{3mm}1. Social activity & 1.272& 1.470 &1.337 & -0.287 & 0.051 & 0.388 & 1.000\\
      \textit{Physical activity}\\
      \hspace{3mm}2. Physical activity &145.829 & 125.573 & 122.270 & 489.123 & 272.545 & 147.046 & 1.000\\
      \textit{Functional health status} \\
      \hspace{3mm}3. Doing household chores & 1.130 & 1.175 & 1.265 & 0.873 & 0.858 & 0.857 & 0.348 \\
      \hspace{3mm}4. Preparing hot meals & 1.134 & 1.168 & 1.218 & 0.868 & 0.840 & 0.795 & 0.361\\
      \hspace{3mm}5. Shopping for groceries &1.144 & 1.163 & 1.217 & 0.843 & 0.793 & 0.739 & 0.407\\
      \hspace{3mm}6. Managing money & 1.243 &1.238 & 1.267 & 0.968 & 0.899 & 0.830 & 0.372 \\
      \hspace{3mm}7. Taking medications &1.091 &1.081 & 1.097 & 0.975 & 0.938 & 0.912 & 0.158\\
      \textit{Depressive symptoms} \\
      \hspace{3mm}8. Bothered by things that usually did not bother me & 2.041 & 1.765 & 1.906 & 1.718 & - & - & 0.660\\
      \hspace{3mm}9. Trouble keeping mind on tasks & 1.926 &1.739 & 1.888 & 1.682 & - & - & 0.600\\
      \hspace{3mm}10. Felt depressed &1.987 &1.759 & 1.896 & 1.673 & - & - & 0.737\\
      \hspace{3mm}11. Felt that everything was an effort & 2.022 &1.822 & 1.896 & 1.721 & - & - & 0.692\\
      \hspace{3mm}12. Felt fearful &1.354 &1.275 & 1.904 & 1.224 & - & - & 0.335\\
      \hspace{3mm}13. Restless sleep &2.038 &2.043 & 1.323 & 1.919 & - & - & 0.455\\
      \hspace{3mm}14. Felt lonely & 1.523 &1.438 & 2.069 & 1.375 & - & - & 0.447\\
      \hspace{3mm}15. Could not get going &1.369 & 1.297 & 1.370 & 1.236 & - & - & 0.389\\
    \hline
\end{tabular}}
\end{table} 

\begin{table}[ht!]
\centering
\caption{Coefficient estimates ($\beta$s and $\pi$s) in the structural model of our strong invariance ASM. 99\% bootstrap confidence intervals are obtained for standardized estimates.  SA stands for social activity, PA physical activity, FHS functional health status, and DS depressive symptoms.}
\bigskip
\label{tab:pathCoef}
\renewcommand{\arraystretch}{1.0}
{\begin{tabular}{lccccc}
      \hline
        Coefficient & Path & Estimate  & Standardized & 99\% CI & $p$-value \\
         &  & & Estimate & for Std Est &  \\
    \hline
       $\beta_{13,1}$ & $SA_1 \rightarrow FHS_1$ & -0.064 & -0.105 & (-0.128, -0.081) & $<0.001$\\
       $\beta_{23,1}$ & $PA_1 \rightarrow FHS_1$ & -0.001 & -0.146 & (-0.172, -0.118) & $<0.001$\\
       $\beta_{14,1}$ & $SA_1 \rightarrow DS_1$ & -0.053 & -0.085 & (-0.110, -0.058) & $<0.001$\\
       $\beta_{24,1}$ & $PA_1 \rightarrow DS_1$ & 0.001 & 0.056 & (0.034, 0.079) & $<0.001$\\
       $\beta_{34,1}$ & $FHS_1 \rightarrow DS_1$ & 0.322 & 0.309 & (0.269, 0.349) & $<0.001$\\
       $\beta_{13,2}$ & $SA_2 \rightarrow FHS_2$ & -0.050 & -0.080 & (-0.100, -0.059) & $<0.001$\\
       $\beta_{23,2}$ & $PA_2 \rightarrow FHS_2$ & -0.001 & -0.081 & (-0.104, -0.058) & $<0.001$\\
       $\beta_{14,2}$ & $SA_2 \rightarrow DS_2$ & -0.008 & -0.015 & (-0.039, 0.010) & $0.119$\\
       $\beta_{24,2}$ & $PA_2 \rightarrow DS_2$ & 0.001 & 0.098 & (0.071, 0.125) & $<0.001$\\
       $\beta_{34,2}$ & $FHS_2 \rightarrow DS_2$ & 0.213 & 0.246 & (0.205, 0.287) & $<0.001$\\
       $\beta_{13,3}$ & $SA_3 \rightarrow FHS_3$ & -0.041 & -0.054 & (-0.072, -0.035) & $<0.001$\\
       $\beta_{23,3}$ & $PA_3 \rightarrow FHS_3$ & -0.001& -0.094 & (-0.116, -0.071) & $<0.001$\\
       $\beta_{14,3}$ & $SA_3 \rightarrow DS_3$ & -0.009 & -0.015 & (-0.038, 0.009) & $0.106$\\
       $\beta_{24,3}$ & $PA_3 \rightarrow DS_3$ & 0.000 & 0.044 & (0.019, 0.070) & $<0.001$\\
       $\beta_{34,3}$ & $FHS_3 \rightarrow DS_3$ & 0.190 & 0.227 & (0.188, 0.265) & $<0.001$\\
      \hline
       $\pi_{1,1 \rightarrow 2}$ & $SA_1 \rightarrow SA_2$ & 0.362 & 0.332 & (0.302, 0.362) & $<0.001$\\
       $\pi_{1,1 \rightarrow 3}$ & $SA_1 \rightarrow SA_3$ & 0.216 & 0.203 & (0.173, 0.233) & $<0.001$\\
       $\pi_{1,2 \rightarrow 3}$ & $SA_2 \rightarrow SA_3$ & 0.313 & 0.321 & (0.290, 0.352) & $<0.001$\\
       $\pi_{2,1 \rightarrow 2}$ & $PA_1 \rightarrow PA_2$ & 0.304 & 0.331 & (0.303, 0.358) & $<0.001$\\
       $\pi_{2,1 \rightarrow 3}$ & $PA_1 \rightarrow PA_3$ & 0.180 & 0.202 & (0.173, 0.232) & $<0.001$\\ 
       $\pi_{2,2 \rightarrow 3}$ & $PA_2 \rightarrow PA_3$ & 0.320 & 0.329 & (0.298, 0.359) & $<0.001$\\    
       $\pi_{3,1 \rightarrow 2}$ & $FHS_1 \rightarrow FHS_2$ & 0.596 & 0.527 & (0.463, 0.591) & $<0.001$\\
       $\pi_{3,1 \rightarrow 3}$ & $FHS_1 \rightarrow FHS_3$ & 0.309 & 0.233 & (0.163, 0.300) & $<0.001$\\
       $\pi_{3,2 \rightarrow 3}$ & $FHS_2 \rightarrow FHS_3$ & 0.576 & 0.493 & (0.420, 0.563) & $<0.001$\\
       $\pi_{4,1 \rightarrow 2}$ & $DS_1 \rightarrow DS_2$ & 0.417 & 0.445 & (0.411, 0.477) & $<0.001$\\
       $\pi_{4,1 \rightarrow 3}$ & $DS_1 \rightarrow DS_3$ & 0.248 & 0.235 & (0.197, 0.271) & $<0.001$\\
       $\pi_{4,2 \rightarrow 3}$ & $DS_2 \rightarrow DS_3$ & 0.420 & 0.373 & (0.335, 0.412) & $<0.001$\\
    \hline
\end{tabular}}
\end{table} 

\begin{table}[ht]
\centering
\caption{Standardized coefficient estimates for the covariates in our strong invariance ASM (denoted by $c$ in Equation~\ref{eqn:order2auto}). Italic coefficients are significant. URB stands for rural/urban, MAR marital status, and EDU educational attainment.}
\bigskip
\label{tab:covariatecoeff}
\renewcommand{\arraystretch}{1.5}
{\begin{tabular}{lccccc}
      \hline
       & & & Covariates && \\
       & SEX & AGE & URB & MAR & EDU \\
    \hline
    Social Activity \\
      ${t=1}$ & \textit{0.041} & \textit{0.046} & \textit{0.077} & 0.005 & \textit{0.133}\\
      ${t=2}$ & \textit{0.028} & -0.020 & \textit{0.078} & \textit{0.039} & \textit{0.111} \\
      ${t=3}$ & \textit{0.029} & \textit{-0.043} & \textit{0.025} & \textit{0.024} & \textit{0.088} \\
      Physical Activity \\
      ${t=1}$ & \textit{-0.174} & \textit{-0.220} & \textit{-0.173} & \textit{-0.038} & \textit{-0.121}\\
      ${t=2}$ & \textit{-0.069} & \textit{-0.180} & \textit{-0.065} & -0.009 & \textit{-0.059} \\
      ${t=3}$ & -0.008 & \textit{-0.077} & \textit{-0.079} & \textit{-0.022} & -0.007 \\
      Functional Health Status \\
      ${t=1}$ & \textit{0.033} & \textit{0.173} & \textit{-0.038} & $0.008$ & \textit{-0.108}\\
      ${t=2}$ & 0.003 & \textit{0.098} & 0.009 & -0.010 & \textit{-0.051} \\
      ${t=3}$ & \textit{0.030} & \textit{0.066} & \textit{-0.035} & 0.010 & -0.006 \\
      Depressive Symptoms \\
      ${t=1}$ & \textit{0.141} & -0.026 & \textit{-0.076} & \textit{0.071} & \textit{-0.082}\\
      ${t=2}$ & \textit{0.076} & \textit{-0.090} & -0.022 & \textit{0.040} & \textit{-0.029} \\
      ${t=3}$ & \textit{0.064} & \textit{-0.028} & \textit{-0.024} & 0.000 & \textit{-0.027} \\
    \hline
\end{tabular}}
\end{table} 

We summarize the coefficient estimates in the structural model of our strong invariance ASM in Table \ref{tab:pathCoef} and Figure \ref{fig:graphCoef}.  Our ASM demonstrates a recurrent mediation relationship: social activity and physical activity independently and consistently affected depressive symptoms by mediating through functional health status (Table \ref{tab:pathCoef}, Figure \ref{fig:graphCoef}). The results show that participants who engaged in social activity and physical activity more frequently perceived lower levels of functional difficulties and fewer depressive symptoms. These structural relationships remained in 2015 even after conditioning on the same relationships measured in 2011 and 2013. 
In addition, we observe a positive relationship between physical activity and depressive symptoms consistently in three waves. The explained variances in depressive symptoms in the ASM gradually increased over time (wave1: $R^2$ = 17.8\%; wave2: $R^2$ = 34.0\%; and wave3: $R^2$ = 43.2\%). 

Furthermore, most covariates have a significant influence on the variables in the ASM (see Table \ref{tab:covariatecoeff}). Specifically, female participants engage in a lower level of PA and a higher level of SA, and report depressive symptoms more frequently than male counterparts. Older participants report less PA and SA and more functional difficulties, but perceive fewer depressive symptoms than younger ones. Urban residents show more participation in SA and less in PA, and fewer functional difficulties and depressive symptoms than rural dwellers. Non-married respondents are more likely to show depressive symptoms than married ones. Last but not least, more educated respondents report less PA, more SA, better functional health and fewer depressive symptoms. On the other hand, the covariates do not affect the estimates of the structural relationships.  

To check whether the parameter estimates are sensible, we calculate the expected values for functional health status and depressive symptoms in 2011, 2013 and 2015, using the estimated coefficients, and compare them with the sample means (Figure \ref{fig:compMean}). The comparison shows that the predicted means are close to the observed mean values with minor deviances.

Our findings suggest that social activity and physical activity may simultaneously and consistently mitigate depressive symptoms by preserving functional capabilities during the aging. However, the concurrence of higher levels of physical activity and elevated depressive symptoms may be explained by negative impacts of domestic and occupational physical activity, because they are major sources of activity among Chinese populations, especially among those with lower socioeconomic status [\cite{chen2012relationships}; \cite{deng2018relationships}]. Our findings indicate the need for long-term monitoring of the socio-behavioral factors for depressive symptoms among Chinese elderly. An important implication from our analysis is that programs and interventions should consider efforts to promote both social and physical activities, which may be beneficial for functional and mental health. Strategies may include emphasizing the role of senior social organizations, encouraging social interactions and improving infrastructure for physical exercises in rural and urban communities.  These interventions may increase the participation of social and leisure-time physical activities among Chinese older adults. Since the structural relationships we inferred here are recurrent, the interventions suggested above will be meaningful for middle-aged and older adults for the future.

\begin{figure}[h!]
\begin{center}
  \includegraphics[width=1.0\textwidth]{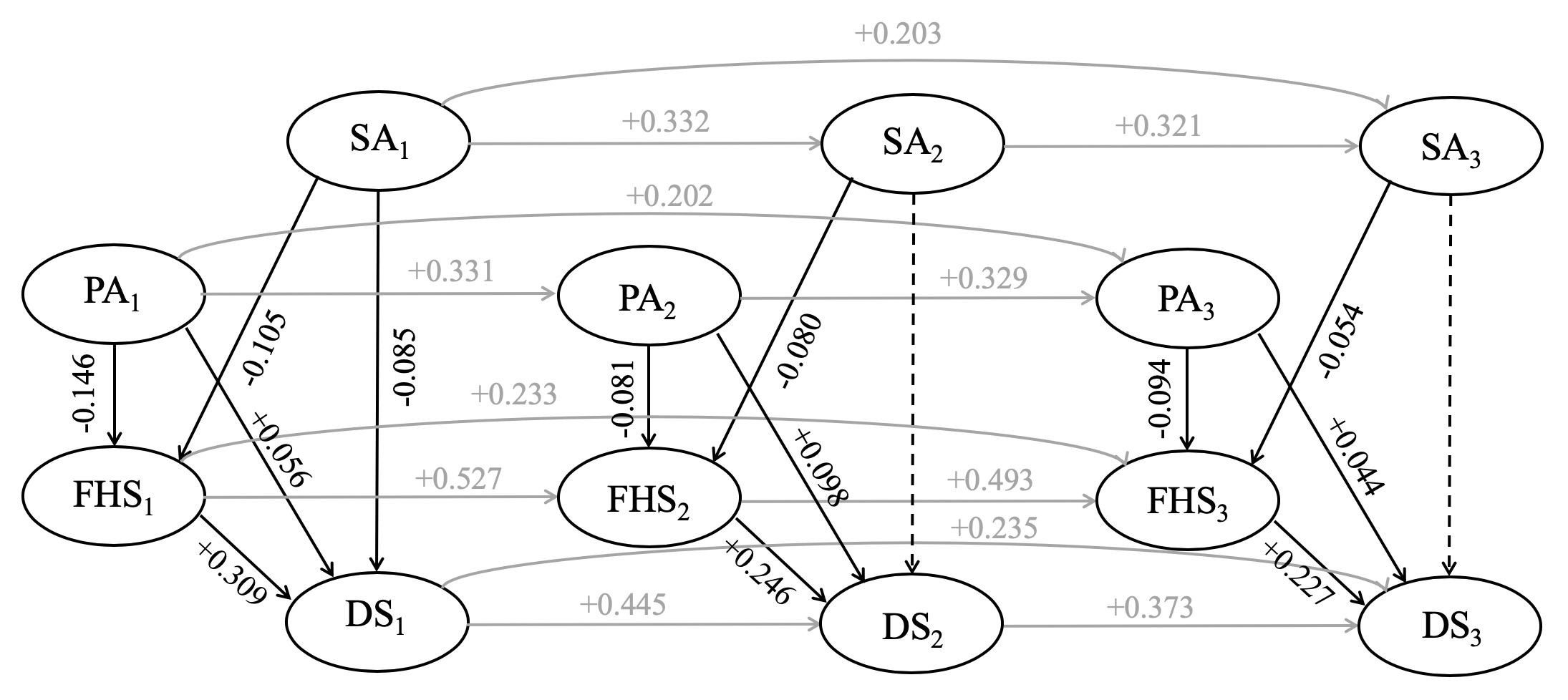}\\
  \caption{Coefficient estimates in the structural model of our order-2 strong invariance ASM over three time points. Black estimates are standardized $\beta$ coefficients, which are the effects among factors, and gray estimates are standardized $\pi$ coefficients, which are the autoregressive effects (see Equation~\ref{eqn:order2auto}). Solid lines indicate statistically significant coefficient estimates, whereas indicate insignificant ones.  SA stands for social activity, PA physical activity, FHS functional health status, and DS depressive symptoms.}
  \label{fig:graphCoef}
\end{center}
\end{figure}

\begin{figure}[h!]
\begin{center}
  \includegraphics[width=1.0\textwidth]{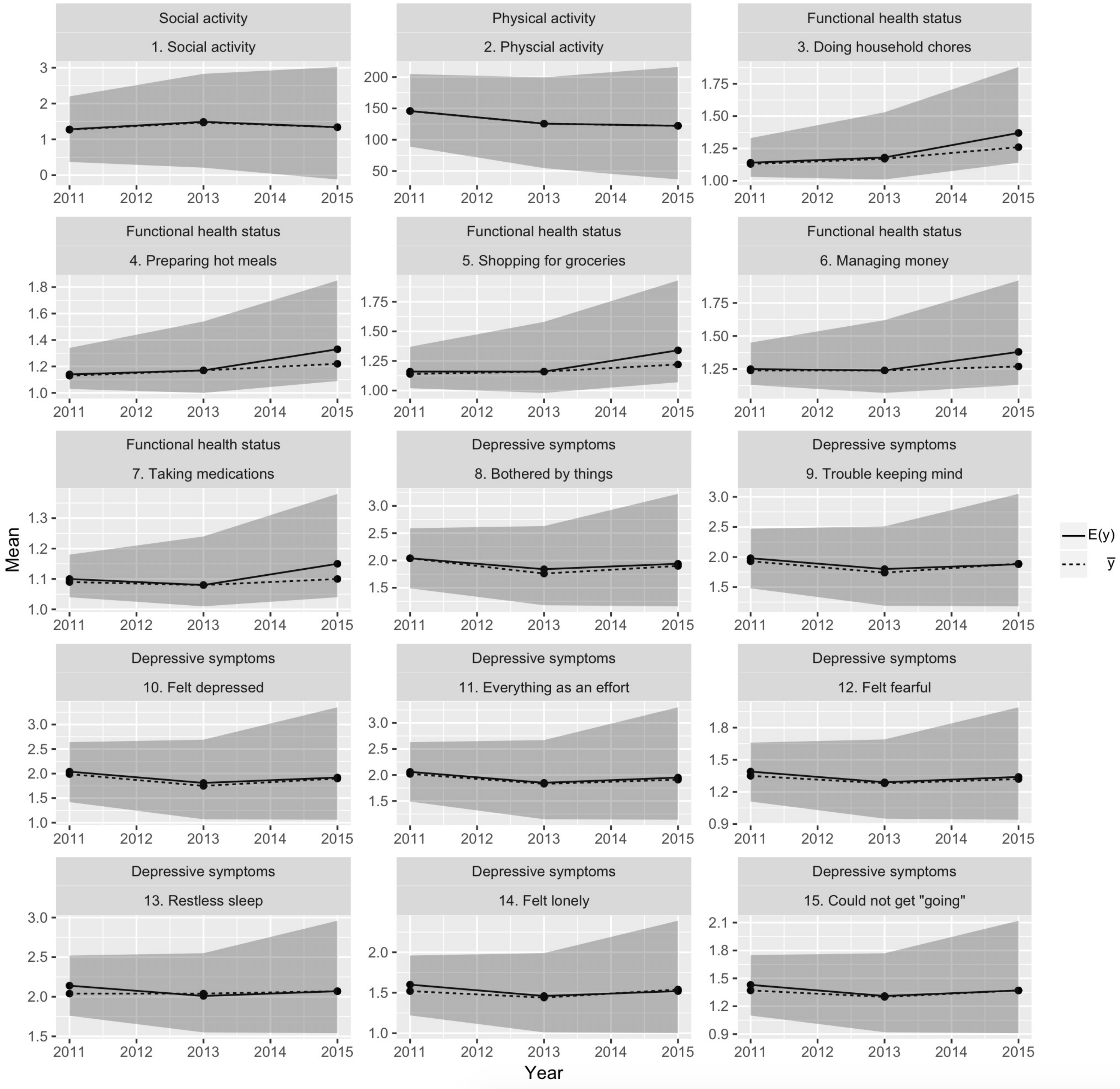}\\
  \caption{Comparison between observed means ($\bar{y}$) and expected values $\mathbb{E}(y)$ from our strong invariance ASM with their 99\% confidence intervals of observed items of social activity, physical activity, functional health status and depressive symptoms.}
  \label{fig:compMean}
\end{center}
\end{figure}

\clearpage

\section{Discussion}
\label{discussion}

In this paper, we propose the ASM as a novel model for investigating the dynamics of a system using multivariate longitudinal data. The ASM extends the previous work by modeling the complex relationships within the template structure over time while accounting for autoregressive dependency. This model has several main features: (i) the structural model that captures complex relationships among multiple variables and the structure can be preserved over multiple time points; (ii) the autoregressive component that accounts for dependency over time; (iii) accounting for covariates and thus reducing bias in parameter estimates; and (iv) allowing for measurement invariance test under the complete model with the three features above.  Applying our ASM to the CHARLS data,  we examine how complex structural relationships evolve over time and show that social activity and physical activity are simultaneously and consistently associated with depressive symptoms by mediating through functional health status over the course of five years.

On the other hand, our ASM has several limitations.  First, we assume a multivariate normal distribution for the observed variables and linear relationships.  Alternatively, one can consider nonparametric structural equations [\cite{pearl2009causality}] and probabilistic graphical model [\cite{koller2009probabilistic}]. Second, accounting for covariates leads to many parameters in the model, which may reduce the stability in estimated parameters.  Additionally, we assume that the impact of covariates is independent of time.  Allowing for time-dependent covariates will include even more parameters.  Techniques, such as the inverse probability of treatment weights [\cite{robins2000marginal}], can help address the challenges of high dimensional covariates and time-dependent confounding. Last but not least, when we investigate a temporal mechanism with extensive repetitions of measurements, the inference techniques of classical SEM may not be optimal to estimate a large number of parameters. Alternative inference methods are needed to handle high-dimensional models. 

In our ASM and its application here, if one variable influences another, we assume that the impact can be measured at the same point without delay. However, some behavioral mechanisms may take effect after a time lag and the optimal time for measuring the effect depends on the underlying process [\cite{selig2009mediation}]. For example, the treatment for a chronic medical condition may take time for the condition to improve. Hence, there is considerable interest in the timing of measurements and how the time lags impact the inference of the structural relationships [\cite{cole2003testing}; \cite{dormann2015optimal}].  In our ongoing work, we will extend our ASM to account for such time lags.  This extended model will provide additional flexibility.

\section*{Acknowledgments}

We thank the CHARLS research and field team and all respondents for their contributions. 


\bibliographystyle{imsart-nameyear}
\bibliography{main}

\begin{thebibliography}{28}

\bibitem[\protect\citeauthoryear{Asparouhov, Hamaker and
  Muth{\'e}n}{2018}]{asparouhov2018dynamic}
\begin{barticle}[author]
\bauthor{\bsnm{Asparouhov},~\bfnm{Tihomir}\binits{T.}},
  \bauthor{\bsnm{Hamaker},~\bfnm{Ellen~L}\binits{E.~L.}} \AND
  \bauthor{\bsnm{Muth{\'e}n},~\bfnm{Bengt}\binits{B.}}
(\byear{2018}).
\btitle{Dynamic structural equation models}.
\bjournal{Structural Equation Modeling: A Multidisciplinary Journal}
\bvolume{25}
\bpages{359--388}.
\end{barticle}
\endbibitem

\bibitem[\protect\citeauthoryear{Banister, Bloom and
  Rosenberg}{2012}]{banister2012population}
\begin{bincollection}[author]
\bauthor{\bsnm{Banister},~\bfnm{Judith}\binits{J.}},
  \bauthor{\bsnm{Bloom},~\bfnm{David~E}\binits{D.~E.}} \AND
  \bauthor{\bsnm{Rosenberg},~\bfnm{Larry}\binits{L.}}
(\byear{2012}).
\btitle{Population aging and economic growth in China}.
In \bbooktitle{the Chinese economy}
\bpages{114--149}.
\bpublisher{Springer}.
\end{bincollection}
\endbibitem

\bibitem[\protect\citeauthoryear{Bollen}{1989}]{Bollen1989Structural}
\begin{bbook}[author]
\bauthor{\bsnm{Bollen},~\bfnm{Kenneth~A}\binits{K.~A.}}
(\byear{1989}).
\btitle{Structural equations with latent variables.}
\bpublisher{New York New York John Wiley \& Sons}.
\end{bbook}
\endbibitem

\bibitem[\protect\citeauthoryear{Bollen and
  Curran}{2004}]{bollen2004autoregressive}
\begin{barticle}[author]
\bauthor{\bsnm{Bollen},~\bfnm{Kenneth~A}\binits{K.~A.}} \AND
  \bauthor{\bsnm{Curran},~\bfnm{Patrick~J}\binits{P.~J.}}
(\byear{2004}).
\btitle{Autoregressive latent trajectory (ALT) models a synthesis of two
  traditions}.
\bjournal{Sociological Methods \& Research}
\bvolume{32}
\bpages{336--383}.
\end{barticle}
\endbibitem

\bibitem[\protect\citeauthoryear{Chen et~al.}{2012}]{chen2012relationships}
\begin{barticle}[author]
\bauthor{\bsnm{Chen},~\bfnm{Li-Jung}\binits{L.-J.}},
  \bauthor{\bsnm{Stevinson},~\bfnm{Clare}\binits{C.}},
  \bauthor{\bsnm{Ku},~\bfnm{Po-Wen}\binits{P.-W.}},
  \bauthor{\bsnm{Chang},~\bfnm{Yu-Kai}\binits{Y.-K.}} \AND
  \bauthor{\bsnm{Chu},~\bfnm{Da-Chen}\binits{D.-C.}}
(\byear{2012}).
\btitle{Relationships of leisure-time and non-leisure-time physical activity
  with depressive symptoms: a population-based study of Taiwanese older
  adults}.
\bjournal{International Journal of Behavioral Nutrition and Physical Activity}
\bvolume{9}
\bpages{28}.
\end{barticle}
\endbibitem

\bibitem[\protect\citeauthoryear{Cole and Maxwell}{2003}]{cole2003testing}
\begin{barticle}[author]
\bauthor{\bsnm{Cole},~\bfnm{David~A}\binits{D.~A.}} \AND
  \bauthor{\bsnm{Maxwell},~\bfnm{Scott~E}\binits{S.~E.}}
(\byear{2003}).
\btitle{Testing mediational models with longitudinal data: questions and tips
  in the use of structural equation modeling.}
\bjournal{Journal of Abnormal Psychology}
\bvolume{112}
\bpages{558}.
\end{barticle}
\endbibitem

\bibitem[\protect\citeauthoryear{Deng and Paul}{2018}]{deng2018relationships}
\begin{barticle}[author]
\bauthor{\bsnm{Deng},~\bfnm{Yazhuo}\binits{Y.}} \AND
  \bauthor{\bsnm{Paul},~\bfnm{David~R}\binits{D.~R.}}
(\byear{2018}).
\btitle{The Relationships Between Depressive Symptoms, Functional Health
  Status, Physical Activity, and the Availability of Recreational Facilities: a
  Rural-Urban Comparison in Middle-Aged and Older Chinese Adults}.
\bjournal{International Journal of Behavioral Medicine}
\bvolume{25}
\bpages{322--330}.
\end{barticle}
\endbibitem

\bibitem[\protect\citeauthoryear{Dormann and
  Griffin}{2015}]{dormann2015optimal}
\begin{barticle}[author]
\bauthor{\bsnm{Dormann},~\bfnm{Christian}\binits{C.}} \AND
  \bauthor{\bsnm{Griffin},~\bfnm{Mark~A}\binits{M.~A.}}
(\byear{2015}).
\btitle{Optimal time lags in panel studies.}
\bjournal{Psychological Methods}
\bvolume{20}
\bpages{489}.
\end{barticle}
\endbibitem

\bibitem[\protect\citeauthoryear{Ferrer and
  McArdle}{2003}]{ferrer2003alternative}
\begin{barticle}[author]
\bauthor{\bsnm{Ferrer},~\bfnm{Emilio}\binits{E.}} \AND
  \bauthor{\bsnm{McArdle},~\bfnm{John}\binits{J.}}
(\byear{2003}).
\btitle{Alternative structural models for multivariate longitudinal data
  analysis}.
\bjournal{Structural Equation Modeling: A Multidisciplinary Journal}
\bvolume{10}
\bpages{493--524}.
\end{barticle}
\endbibitem

\bibitem[\protect\citeauthoryear{Fiske, Wetherell and
  Gatz}{2009}]{fiske2009depression}
\begin{barticle}[author]
\bauthor{\bsnm{Fiske},~\bfnm{Amy}\binits{A.}},
  \bauthor{\bsnm{Wetherell},~\bfnm{Julie~Loebach}\binits{J.~L.}} \AND
  \bauthor{\bsnm{Gatz},~\bfnm{Margaret}\binits{M.}}
(\byear{2009}).
\btitle{Depression in older adults}.
\bjournal{Annual Review of Clinical Psychology}
\bvolume{5}
\bpages{363--389}.
\end{barticle}
\endbibitem

\bibitem[\protect\citeauthoryear{Fried et~al.}{2004}]{fried2004social}
\begin{barticle}[author]
\bauthor{\bsnm{Fried},~\bfnm{Linda~P}\binits{L.~P.}},
  \bauthor{\bsnm{Carlson},~\bfnm{Michelle~C}\binits{M.~C.}},
  \bauthor{\bsnm{Freedman},~\bfnm{Marc}\binits{M.}},
  \bauthor{\bsnm{Frick},~\bfnm{Kevin~D}\binits{K.~D.}},
  \bauthor{\bsnm{Glass},~\bfnm{Thomas~A}\binits{T.~A.}},
  \bauthor{\bsnm{Hill},~\bfnm{Joel}\binits{J.}},
  \bauthor{\bsnm{McGill},~\bfnm{Sylvia}\binits{S.}},
  \bauthor{\bsnm{Rebok},~\bfnm{George~W}\binits{G.~W.}},
  \bauthor{\bsnm{Seeman},~\bfnm{Teresa}\binits{T.}},
  \bauthor{\bsnm{Tielsch},~\bfnm{James}\binits{J.}} \betal{et~al.}
(\byear{2004}).
\btitle{A Social Model for Health Promotion for an Aging Population: Initial
  Evidence on the Experience Corps Model}.
\bjournal{journal of Urban Health: Bulletin of the New York Academy of
  Medicine}
\bvolume{81}
\bpages{64}.
\end{barticle}
\endbibitem

\bibitem[\protect\citeauthoryear{Grimm, Ram and
  Estabrook}{2016}]{grimm2016growth}
\begin{bbook}[author]
\bauthor{\bsnm{Grimm},~\bfnm{Kevin~J}\binits{K.~J.}},
  \bauthor{\bsnm{Ram},~\bfnm{Nilam}\binits{N.}} \AND
  \bauthor{\bsnm{Estabrook},~\bfnm{Ryne}\binits{R.}}
(\byear{2016}).
\btitle{Growth modeling: Structural equation and multilevel modeling
  approaches}.
\bpublisher{Guilford Publications}.
\end{bbook}
\endbibitem

\bibitem[\protect\citeauthoryear{Hu and Bentler}{1999}]{hu1999cutoff}
\begin{barticle}[author]
\bauthor{\bsnm{Hu},~\bfnm{Li-tze}\binits{L.-t.}} \AND
  \bauthor{\bsnm{Bentler},~\bfnm{Peter~M}\binits{P.~M.}}
(\byear{1999}).
\btitle{Cutoff criteria for fit indexes in covariance structure analysis:
  Conventional criteria versus new alternatives}.
\bjournal{Structural Equation Modeling: a Multidisciplinary Journal}
\bvolume{6}
\bpages{1--55}.
\end{barticle}
\endbibitem

\bibitem[\protect\citeauthoryear{Koller, Friedman and
  Bach}{2009}]{koller2009probabilistic}
\begin{bbook}[author]
\bauthor{\bsnm{Koller},~\bfnm{Daphne}\binits{D.}},
  \bauthor{\bsnm{Friedman},~\bfnm{Nir}\binits{N.}} \AND
  \bauthor{\bsnm{Bach},~\bfnm{Francis}\binits{F.}}
(\byear{2009}).
\btitle{Probabilistic graphical models: principles and techniques}.
\bpublisher{MIT press}.
\end{bbook}
\endbibitem

\bibitem[\protect\citeauthoryear{Little}{2013}]{little2013longitudinal}
\begin{bbook}[author]
\bauthor{\bsnm{Little},~\bfnm{Todd~D}\binits{T.~D.}}
(\byear{2013}).
\btitle{Longitudinal structural equation modeling}.
\bpublisher{Guilford press}.
\end{bbook}
\endbibitem

\bibitem[\protect\citeauthoryear{MacKinnon}{2008}]{mackinnon2008introduction}
\begin{bbook}[author]
\bauthor{\bsnm{MacKinnon},~\bfnm{David}\binits{D.}}
(\byear{2008}).
\btitle{Introduction to statistical mediation analysis}.
\bpublisher{Routledge}.
\end{bbook}
\endbibitem

\bibitem[\protect\citeauthoryear{Marsh}{1993}]{marsh1993stability}
\begin{barticle}[author]
\bauthor{\bsnm{Marsh},~\bfnm{Herbert~W}\binits{H.~W.}}
(\byear{1993}).
\btitle{Stability of individual differences in multiwave panel studies:
  Comparison of simplex models and one-factor models}.
\bjournal{Journal of Educational Measurement}
\bvolume{30}
\bpages{157--183}.
\end{barticle}
\endbibitem

\bibitem[\protect\citeauthoryear{Mayer}{1986}]{mayer1986cross}
\begin{barticle}[author]
\bauthor{\bsnm{Mayer},~\bfnm{Lawrence~S}\binits{L.~S.}}
(\byear{1986}).
\btitle{On cross-lagged panel models with serially correlated errors}.
\bjournal{Journal of Business \& Economic Statistics}
\bvolume{4}
\bpages{347--357}.
\end{barticle}
\endbibitem

\bibitem[\protect\citeauthoryear{McArdle and
  Nesselroade}{2014}]{mcardle2014longitudinal}
\begin{bbook}[author]
\bauthor{\bsnm{McArdle},~\bfnm{John~J}\binits{J.~J.}} \AND
  \bauthor{\bsnm{Nesselroade},~\bfnm{John~R}\binits{J.~R.}}
(\byear{2014}).
\btitle{Longitudinal data analysis using structural equation models.}
\bpublisher{American Psychological Association}.
\end{bbook}
\endbibitem

\bibitem[\protect\citeauthoryear{Millsap and
  Cham}{2012}]{millsap2012investigating}
\begin{bbook}[author]
\bauthor{\bsnm{Millsap},~\bfnm{Roger~E}\binits{R.~E.}} \AND
  \bauthor{\bsnm{Cham},~\bfnm{Heining}\binits{H.}}
(\byear{2012}).
\btitle{Investigating factorial invariance in longitudinal data.}
\bpublisher{Guilford Press}.
\end{bbook}
\endbibitem

\bibitem[\protect\citeauthoryear{Muth{\'e}n and
  Muth{\'e}n}{2017}]{muthen2017mplus}
\begin{bbook}[author]
\bauthor{\bsnm{Muth{\'e}n},~\bfnm{LK}\binits{L.}} \AND
  \bauthor{\bsnm{Muth{\'e}n},~\bfnm{B}\binits{B.}}
(\byear{2017}).
\btitle{Mplus \text{User's Guide, 8th ed}}.
\bpublisher{Muth{\'e}n \& Muth{\'e}n.}
\end{bbook}
\endbibitem

\bibitem[\protect\citeauthoryear{Pearl}{2009}]{pearl2009causality}
\begin{bbook}[author]
\bauthor{\bsnm{Pearl},~\bfnm{Judea}\binits{J.}}
(\byear{2009}).
\btitle{Causality}.
\bpublisher{Cambridge university press}.
\end{bbook}
\endbibitem

\bibitem[\protect\citeauthoryear{Robins, Hern{\'a}n and
  Brumback}{2000}]{robins2000marginal}
\begin{barticle}[author]
\bauthor{\bsnm{Robins},~\bfnm{James~M}\binits{J.~M.}},
  \bauthor{\bsnm{Hern{\'a}n},~\bfnm{Miguel~Angel}\binits{M.~A.}} \AND
  \bauthor{\bsnm{Brumback},~\bfnm{Babette}\binits{B.}}
(\byear{2000}).
\btitle{Marginal Structural Models and Causal Inference in Epidemiology}.
\bjournal{Epidemiology}
\bvolume{11}
\bpages{551}.
\end{barticle}
\endbibitem

\bibitem[\protect\citeauthoryear{Selig and Preacher}{2009}]{selig2009mediation}
\begin{barticle}[author]
\bauthor{\bsnm{Selig},~\bfnm{James~P}\binits{J.~P.}} \AND
  \bauthor{\bsnm{Preacher},~\bfnm{Kristopher~J}\binits{K.~J.}}
(\byear{2009}).
\btitle{Mediation models for longitudinal data in developmental research}.
\bjournal{Research in Human Development}
\bvolume{6}
\bpages{144--164}.
\end{barticle}
\endbibitem

\bibitem[\protect\citeauthoryear{Widaman, Ferrer and
  Conger}{2010}]{widaman2010factorial}
\begin{barticle}[author]
\bauthor{\bsnm{Widaman},~\bfnm{Keith~F}\binits{K.~F.}},
  \bauthor{\bsnm{Ferrer},~\bfnm{Emilio}\binits{E.}} \AND
  \bauthor{\bsnm{Conger},~\bfnm{Rand~D}\binits{R.~D.}}
(\byear{2010}).
\btitle{Factorial invariance within longitudinal structural equation models:
  Measuring the same construct across time}.
\bjournal{Child Development Perspectives}
\bvolume{4}
\bpages{10--18}.
\end{barticle}
\endbibitem

\bibitem[\protect\citeauthoryear{Xu et~al.}{2016}]{xu2016decomposing}
\begin{barticle}[author]
\bauthor{\bsnm{Xu},~\bfnm{Yongjian}\binits{Y.}},
  \bauthor{\bsnm{Yang},~\bfnm{Jinjuan}\binits{J.}},
  \bauthor{\bsnm{Gao},~\bfnm{Jianmin}\binits{J.}},
  \bauthor{\bsnm{Zhou},~\bfnm{Zhongliang}\binits{Z.}},
  \bauthor{\bsnm{Zhang},~\bfnm{Tao}\binits{T.}},
  \bauthor{\bsnm{Ren},~\bfnm{Jianping}\binits{J.}},
  \bauthor{\bsnm{Li},~\bfnm{Yanli}\binits{Y.}},
  \bauthor{\bsnm{Qian},~\bfnm{Yuyan}\binits{Y.}},
  \bauthor{\bsnm{Lai},~\bfnm{Sha}\binits{S.}} \AND
  \bauthor{\bsnm{Chen},~\bfnm{Gang}\binits{G.}}
(\byear{2016}).
\btitle{Decomposing socioeconomic inequalities in depressive symptoms among the
  elderly in China}.
\bjournal{BMC Public Health}
\bvolume{16}
\bpages{1214}.
\end{barticle}
\endbibitem

\bibitem[\protect\citeauthoryear{Zhao et~al.}{2012}]{zhao2012cohort}
\begin{barticle}[author]
\bauthor{\bsnm{Zhao},~\bfnm{Yaohui}\binits{Y.}},
  \bauthor{\bsnm{Hu},~\bfnm{Yisong}\binits{Y.}},
  \bauthor{\bsnm{Smith},~\bfnm{James~P}\binits{J.~P.}},
  \bauthor{\bsnm{Strauss},~\bfnm{John}\binits{J.}} \AND
  \bauthor{\bsnm{Yang},~\bfnm{Gonghuan}\binits{G.}}
(\byear{2012}).
\btitle{Cohort profile: The China health and retirement longitudinal study
  (CHARLS)}.
\bjournal{International Journal of Epidemiology}
\bvolume{43}
\bpages{61--68}.
\end{barticle}
\endbibitem

\bibitem[\protect\citeauthoryear{Zhou et~al.}{2019}]{zhou2019mortality}
\begin{barticle}[author]
\bauthor{\bsnm{Zhou},~\bfnm{Maigeng}\binits{M.}},
  \bauthor{\bsnm{Wang},~\bfnm{Haidong}\binits{H.}},
  \bauthor{\bsnm{Zeng},~\bfnm{Xinying}\binits{X.}},
  \bauthor{\bsnm{Yin},~\bfnm{Peng}\binits{P.}},
  \bauthor{\bsnm{Zhu},~\bfnm{Jun}\binits{J.}},
  \bauthor{\bsnm{Chen},~\bfnm{Wanqing}\binits{W.}},
  \bauthor{\bsnm{Li},~\bfnm{Xiaohong}\binits{X.}},
  \bauthor{\bsnm{Wang},~\bfnm{Lijun}\binits{L.}},
  \bauthor{\bsnm{Wang},~\bfnm{Limin}\binits{L.}},
  \bauthor{\bsnm{Liu},~\bfnm{Yunning}\binits{Y.}} \betal{et~al.}
(\byear{2019}).
\btitle{Mortality, morbidity, and risk factors in China and its provinces,
  1990--2017: a systematic analysis for the Global Burden of Disease Study
  2017}.
\bjournal{The Lancet}.
\end{barticle}
\endbibitem

\end{thebibliography}


\end{document}